\documentclass[10pt, conference]{IEEEtran}
\usepackage[left=1.9cm, right=1.9cm, top=0.9cm, bottom=2.cm]{geometry}
\usepackage{graphics, subfigure, color, epsfig, psfrag, amsmath, amsfonts, amssymb, eucal, balance, url}
\newtheorem{lem}{Lemma}
\newtheorem{prop}{Proposition}
\newtheorem{defn}{Definition}

\DeclareMathAlphabet{\eurm}{U}{eur}{m}{n}
\DeclareMathAlphabet{\mathbsf}{OT1}{cmss}{bx}{n}
\DeclareMathAlphabet{\mathssf}{OT1}{cmss}{m}{sl}
\DeclareMathAlphabet{\mathcsf}{OT1}{cmss}{sbc}{n}

\newcommand{\randomvalue}[1]{\eurm{\uppercase{#1}}}

\DeclareSymbolFont{bsfletters}{OT1}{cmss}{bx}{n}  
\DeclareSymbolFont{ssfletters}{OT1}{cmss}{m}{n}
\DeclareMathSymbol{\bsfGamma}{0}{bsfletters}{'000}
\DeclareMathSymbol{\ssfGamma}{0}{ssfletters}{'000}
\DeclareMathSymbol{\bsfDelta}{0}{bsfletters}{'001}
\DeclareMathSymbol{\ssfDelta}{0}{ssfletters}{'001}
\DeclareMathSymbol{\bsfTheta}{0}{bsfletters}{'002}
\DeclareMathSymbol{\ssfTheta}{0}{ssfletters}{'002}
\DeclareMathSymbol{\bsfLambda}{0}{bsfletters}{'003}
\DeclareMathSymbol{\ssfLambda}{0}{ssfletters}{'003}
\DeclareMathSymbol{\bsfXi}{0}{bsfletters}{'004}
\DeclareMathSymbol{\ssfXi}{0}{ssfletters}{'004}
\DeclareMathSymbol{\bsfPi}{0}{bsfletters}{'005}
\DeclareMathSymbol{\ssfPi}{0}{ssfletters}{'005}
\DeclareMathSymbol{\bsfSigma}{0}{bsfletters}{'006}
\DeclareMathSymbol{\ssfSigma}{0}{ssfletters}{'006}
\DeclareMathSymbol{\bsfUpsilon}{0}{bsfletters}{'007}
\DeclareMathSymbol{\ssfUpsilon}{0}{ssfletters}{'007}
\DeclareMathSymbol{\bsfPhi}{0}{bsfletters}{'010}
\DeclareMathSymbol{\ssfPhi}{0}{ssfletters}{'010}
\DeclareMathSymbol{\bsfPsi}{0}{bsfletters}{'011}
\DeclareMathSymbol{\ssfPsi}{0}{ssfletters}{'011}
\DeclareMathSymbol{\bsfOmega}{0}{bsfletters}{'012}
\DeclareMathSymbol{\ssfOmega}{0}{ssfletters}{'012}

\newcommand{\rva}{{\randomvalue{a}}}	

\newcommand{\rve}{{\randomvalue{e}}}	

\newcommand{\rvs}{{\randomvalue{s}}}	

\newcommand{\rvu}{{\randomvalue{u}}}	

\newcommand{\rvw}{{\randomvalue{w}}}	

\newcommand{\rvx}{{\randomvalue{x}}}	

\newcommand{\rvy}{{\randomvalue{y}}}	

\newcommand{\rvz}{{\randomvalue{z}}}	

\begin{document}

\title{On OR Many-Access Channels}
\author{\IEEEauthorblockN{Wenyi Zhang, Lingyan Huang}
\IEEEauthorblockA{University of Science and Technology of China\\
Email: wenyizha@ustc.edu.cn}
}

\maketitle

\begin{abstract}
OR multi-access channel is a simple model where the channel output is the Boolean OR among the Boolean channel inputs. We revisit this model, showing that employing Bloom filter, a randomized data structure, as channel inputs achieves its capacity region with joint decoding and the symmetric sum rate of $\ln 2$ bits per channel use without joint decoding. We then proceed to the ``many-access'' regime where the number of potential users grows without bound, treating both activity recognition and message transmission problems, establishing scaling laws which are optimal within a constant factor, based on Bloom filter channel inputs.
\end{abstract}

\section{Introduction}
\label{sec:intro}

Motivated by the need of massive connectivity in future wireless networks, it is of considerable interest to investigate multi-access systems where there are an exceedingly large number of potential users, among which a small fraction of active ones spontaneously attempt to send information. In order to extract its fundamental feature of massiveness, such a setting is distinct from the classical multi-access channel (MAC) model in the following aspects: (1) the total user population size is increasing, (2) the average active user population size is also increasing but with its fraction in the total user population vanishing, and (3) the data packet size per user is fixed (usually small) or increasing depending upon the total user population size. The central question therefore is how the transmitted signal duration should scale with the user population size for reliable communication.

The case of Gaussian MAC in the aforementioned regime has been treated in \cite{chen16:arxiv} with the name of ``many access'' explicitly proposed therein. In this work, we consider the case of OR MAC, where the channel output is the Boolean OR among the Boolean channel inputs. As will be seen, on the one hand, the OR MAC is a deceptively simple model, when one realizes that time-sharing achieves every point in its capacity region; on the other hand, studying the OR MAC can still shed light on building efficient many-access systems. We propose coding schemes for OR MAC and extend them to the many-access regime, using a randomized data structure called Bloom filter. For OR MAC, the capacity region and the symmetric sum rate of $\ln 2$ bits per channel use can be achieved, with and without joint decoding respectively, using Bloom filter channel inputs. In the many-access regime, both activity recognition and message transmission problems are considered, and scaling laws are established based on Bloom filter channel inputs. Unlike Gaussian many-access channels \cite{chen16:arxiv}, where sharp characterizations of optimal scaling laws have been established utilizing tools from sparse recovery and a two-phase scheme that separates activity recognition and message transmission has been shown to be asymptotically optimal, here for OR many-access channels, our study can only establish scaling laws which are optimal within a constant factor, and our coding scheme suggests that in the activity recognition phase it may be beneficial to leave some ambiguity about the active user population to resolve in the message transmission phase.

The remaining part of this paper is organized as follows. Section \ref{sec:or-mac} and Section \ref{sec:bloom-filter} introduce OR MAC and Bloom filter respectively. Section \ref{sec:bf-or-mac} then revisits the OR MAC under Bloom filter channel inputs, and Section \ref{sec:bf-or-many} treats the many-access regime.

\section{Preliminary of OR MAC}
\label{sec:or-mac}

The OR MAC is a memoryless noiseless MAC as
\begin{eqnarray}
\rvy \! = \! \rvx_1 \! \vee \! \rvx_2 \! \vee \! \ldots \! \vee \! \rvx_N, \;\;\;\;\rvx_n, n = 1, \ldots, N, \rvy \in \{0, 1\}.
\end{eqnarray}
So the channel output is ``0'' if and only if all the channel inputs are ``0''s, and is ``1'' if at least one of the channel inputs is ``1''. The OR MAC is one of the simplest toy examples in multiuser information theory \cite[Example 15.3.2]{cover06:book}.\footnote{Therein the example is in form of binary multiplier channels, equivalent to the OR MAC with $N = 2$.}

As a practical motivation, consider a multi-access system where each user adopts on-off signaling and the receiver front-end is an envelope detector. When the noise level is negligibly low, the input-output relationship is described by the OR MAC \cite{cohen71:tcom}. Note that when the number of users is large and most of them do not send anything at all, such a multi-access system can be attractive since it does not require coherent signal processing at the receiver.

The capacity region of the $N$-user OR MAC is simply
\begin{eqnarray}
\mathcal{C}_N = \left\{\underline{R}: R_1 + R_2 + \ldots + R_N \leq 1 \;\mbox{bit/c.u.}\right\},
\end{eqnarray}
where c.u. stands for ``channel use''. The converse of $\mathcal{C}_N$ is due to that $R_1 + R_2 + \ldots + R_N \leq I(\rvx_1, \rvx_2, \ldots, \rvx_N; \rvy) \leq H(\rvy) \leq 1$ bit/c.u.. The achievability of $\mathcal{C}_N$ can be shown via time-sharing; that is, to achieve $\underline{R} \in \mathcal{C}_N$, split the channel uses so that user $n$ is allocated a fraction of $R_n$ of the channel uses exclusively.

A time-sharing scheme requires some level of coordination among users, which may not be available. However, an interesting fact is that the capacity region $\mathcal{C}_N$ can also be achieved without time-sharing. For example, in order to achieve the symmetric point of $R_n = 1/N$ bits/c.u., $n = 1, \ldots, N$, we let $\rvx_n \sim \mbox{Bernoulli}(1 - 2^{-1/N})$, $n = 1, \ldots, N$, and perform joint decoding at the receiver.

Without joint decoding, user $n$ achieves a rate of $R_n = I(\rvx_n; \rvy)$, and the sum rate is $R_\mathrm{sum} = \sum_{n = 1}^N R_n$. Under $\rvx_n \sim \mbox{Bernoulli}(1 - 2^{-1/N})$, $n = 1, \ldots, N$, $R_\mathrm{sum}$ quickly tends to a limit of $\ln 2 \approx 0.69$ bits/c.u. with $N$ \cite{hui83:dissertation}. It is worth noting that the loss due to not employing joint decoding is only 31\%; in contrast, such loss is unbounded with $N$ in Gaussian MAC.

\section{Preliminary of Bloom Filters}
\label{sec:bloom-filter}

In this section, we briefly introduce the idea of Bloom filter, which was named after Bloom \cite{bloom70:cacm}.\footnote{For its numerous applications and many variants in computer systems, see, e.g., \url{https://en.wikipedia.org/wiki/Bloom_filter}. Essentially the same idea was described by Mooers in his invention of information retrieval machines in the 1940-50s \cite{mooers54:it}.} A Bloom filter of parameters $(L, K)$, denoted by $\mathsf{BF}(L, K)$, is a length-$L$ array, generated according to the following rule:
\begin{itemize}
\item Initially all the positions of the array are ``0''s.
\item Each of $K$ hash functions independently and uniformly randomly selects one of the positions to set it to ``1''.
\end{itemize}
Note that a position in a Bloom filter is set to ``1'' if it is hashed by at least one of the $K$ hash functions, and that a position may be set to ``1'' by different hash functions several times. Also note that a Bloom filter is not a collection of $L$ mutually independent Bernoulli random variables.

Bloom filter provides a way of storing/retrieving items efficiently. Consider a universe of items, a few among which are to be stored. Let each item in the universe be associated with a Bloom filter of parameters $(L, K)$, independently of all others'. Start with an empty (i.e., all-``0'') length-$L$ array. To store an item, ``superpose'' the Bloom filter of this item on the array; that is, mark a position in the array as ``1'' if this position is ``1'' in the Bloom filter of this item. Repeat this procedure until all items of interest have been stored.

When verifying whether an item has been stored in an array, simply check whether the Bloom filter of this item is ``contained'' in the array (i.e., array containing all ``1''s of the Bloom filter). A remarkable property of the method is that there is no miss; --- if an item has been stored, it will surely be checked out. Though there may be false alarms, it is possible to control the false alarm rate by appropriately choosing the parameters $L$ and $K$; see, e.g., \cite{mitzenmacher02:ton}.

The following three properties of Bloom filters are instrumental for our subsequent analysis.

\begin{lem}
\label{lem:superposition}
({\bf Superposition property}) After superposing two Bloom filters, $\mathsf{BF}(L, K_1)$ and $\mathsf{BF}(L, K_2)$, together, the resulting array is a Bloom filter of parameters $(L, K_1 + K_2)$. That is, we can define a superposition operator ``+'' as
\begin{eqnarray}
\mathsf{BF}(L, K_1) + \mathsf{BF}(L, K_2) = \mathsf{BF}(L, K_1 + K_2).
\end{eqnarray}
\end{lem}

\begin{lem}
\label{lem:uniformity}
({\bf Conditional uniformity property}) Define the weight $\rvw$ of a Bloom filter $\underline{\rvy} = \mathsf{BF}(L, K)$ as the number of ``1''s in $\underline{\rvy}$. Conditioned upon $\rvw$, $\underline{\rvy}$ is uniformly distributed among all its $L \choose \rvw$ possibilities.
\end{lem}

For $\mathsf{BF}(L, K)$, the distribution of $\rvw$ is given by $\mathrm{Pr}[\rvw = w] = w! S(K, w)/L^K$, where $S(K, w)$ is the Stirling number of the second kind, counting the number of ways to partition a set of $K$ elements into $w$ nonempty subsets. But for our analysis the following asymptotic behavior suffices.
\begin{lem}
\label{lem:occupancy}
({\bf Occupancy concentration property}) The number of ``0''s $\rvz = L - \rvw$ in $\mathsf{BF}(L, K)$ satisfies for any $\epsilon > 0$,
\begin{eqnarray}
\label{eqn:occupancy-basic}
\mathrm{Pr}[|\rvz - p L| > \epsilon L] < 2 \exp\left(- \frac{\epsilon^2 L^2}{2 K}\right),
\end{eqnarray}
where $p = (1 - 1/L)^K$. The number of ``0''s $\rvz$ in $\mathsf{BF}(L, K_1 + K_2) = \mathsf{BF}(L, K_1) + \mathsf{BF}(L, K_2)$, conditioned upon $\mathsf{BF}(L, K_1)$, satisfies for any $\epsilon > 0$,
\begin{eqnarray}
\label{eqn:occupancy-conditional}
\mathrm{Pr}\left[|\rvz - p_2 \rvz_1| > \epsilon L\big |\mathsf{BF}(L, K_1)\right] < 2 \exp\left(- \frac{\epsilon^2 L^2}{2 K_2}\right),
\end{eqnarray}
where $\rvz_i$ is the number of ``0''s in $\mathsf{BF}(L, K_i)$, and $p_i = (1 - 1/L)^{K_i}$, $i = 1, 2$.
\end{lem}

Lemmas \ref{lem:superposition} and \ref{lem:uniformity} follow from the rule of generating Bloom filters. The bound (\ref{eqn:occupancy-basic}) in Lemma \ref{lem:occupancy} has been proved in \cite{motwani95:book}, as an exercise of Azuma's inequality. Consider the construction of $\mathsf{BF}(L, K)$, one hash function at a time, progressively. Initially, we have an empty array, and the expected number of ``0''s in the final $\mathsf{BF}(L, K)$ is $\bar{\rvz}_0 = (1 - 1/L)^K L$. After $k$ hash functions, denote the conditional expected number of ``0''s in the final $\mathsf{BF}(L, K)$ by $\bar{\rvz}_k$, $k = 1, 2, \ldots, K$. We have that $\bar{\rvz}_0, \bar{\rvz}_1, \ldots, \bar{\rvz}_K = \rvz$ form a martingale sequence with stepwise absolute difference at most one. Thus both bounds in Lemma \ref{lem:occupancy} follow from Azuma's inequality.

\section{OR MAC Revisited}
\label{sec:bf-or-mac}

Return to the $N$-user OR MAC with a fixed $N$. Our first result is the following:
\begin{prop}
\label{prop:cap-achieving}
Bloom filters as channel inputs achieve the capacity region of the $N$-user OR MAC.
\end{prop}
{\it Outline of Proof:} Let the channel input $\underline{\rvx}_n$ of user $n$ be $\mathsf{BF}(L, K_n)$, for $n = 1, 2, \ldots, N$. The corresponding channel output is $\underline{\rvy}$. We calculate the (normalized) mutual information $(1/L)\cdot I(\underline{\rvx}_\mathcal{S}; \underline{\rvy}|\underline{\rvx}_{\bar{\mathcal{S}}})$, for any subset $\mathcal{S} \subseteq \{1, 2, \ldots, N\}$ and $\bar{\mathcal{S}} = \{1, 2, \ldots, N\} \setminus \mathcal{S}$. Here for simplicity we treat the case of $N = 2$, and the case of general $N$ can be treated analogously.

With $N = 2$, we consider the following:
\begin{eqnarray}
L (R_1 + R_2) < I(\underline{\rvx}_1, \underline{\rvx}_2; \underline{\rvy}) &=& H(\underline{\rvy}),\\
L R_1 < I(\underline{\rvx}_1; \underline{\rvy} | \underline{\rvx}_2) &=& H(\underline{\rvy} | \underline{\rvx}_2),\\
L R_2 < I(\underline{\rvx}_2; \underline{\rvy} | \underline{\rvx}_1) &=& H(\underline{\rvy} | \underline{\rvx}_1).
\end{eqnarray}

To evaluate $H(\underline{\rvy})$, $H(\underline{\rvy}|\underline{\rvx}_2)$ and $H(\underline{\rvy}|\underline{\rvx}_1)$, we need the following result.
\begin{lem}
\label{lem:bf-entropy}
Assuming $\lim_{L \rightarrow \infty} K/L = \kappa > 0$, the (normalized) entropy of $\mathsf{BF}(L, K)$ satisfies
\begin{eqnarray}
\label{eqn:bf-entropy}
\lim_{L \rightarrow \infty} (1/L) \cdot H(\mathsf{BF}(L, K)) = h_2(p),
\end{eqnarray}
where $p = \exp(-\kappa)$ and $h_2(x) = - x \log x - (1 - x) \log (1 - x)$. Assuming $\lim_{L \rightarrow \infty} K_i/L = \kappa_i > 0$, $i = 1, 2$, the (normalized) conditional entropy of $\mathsf{BF}(L, K_1) + \mathsf{BF}(L, K_2)$ conditioned upon $\mathsf{BF}(L, K_1)$ satisfies
\begin{eqnarray}
\label{eqn:bf-entropy-conditional}
&&\lim_{L \rightarrow \infty} (1/L)\cdot H(\mathsf{BF}(L, K_1) + \mathsf{BF}(L, K_2) | \mathsf{BF}(L, K_1))\nonumber\\
&=& p_1 h_2(p_2),
\end{eqnarray}
where $p_i = \exp(-\kappa_i)$, $i = 1, 2$.
\end{lem}

Applying Lemma \ref{lem:bf-entropy}, with $K_i = \kappa_i L$, $i = 1, 2$, we have
\begin{eqnarray}
\label{eqn:hy}
\!\!\!\!\lim_{L \rightarrow \infty} (1/L)\cdot H(\underline{\rvy}) &=& h_2(\exp[-(\kappa_1 + \kappa_2)]),\\
\label{eqn:hyx2}
\!\!\!\!\lim_{L \rightarrow \infty} (1/L)\cdot H(\underline{\rvy}|\underline{\rvx}_2) &=& \exp(-\kappa_2) h_2(\exp(-\kappa_1)),\\
\label{eqn:hyx1}
\!\!\!\!\lim_{L \rightarrow \infty} (1/L)\cdot H(\underline{\rvy}|\underline{\rvx}_1) &=& \exp(-\kappa_1) h_2(\exp(-\kappa_2)).
\end{eqnarray}
By varying $\kappa_1, \kappa_2 > 0$ while keeping $\kappa_1 + \kappa_2 = \ln 2$, from (\ref{eqn:hy}), (\ref{eqn:hyx2}) and (\ref{eqn:hyx1}), we can achieve the capacity region $\mathcal{C}_2 = \{\underline{R}: R_1 + R_2 \leq 1 \;\mbox{bit}\}$. $\Box$

\noindent{\it Outline of Proof of Lemma \ref{lem:bf-entropy}:} Denote the weight of $\mathsf{BF}(L, K)$ by $\rvw$. We have
\begin{eqnarray}
H(\mathsf{BF}(L, K)) &=& H(\mathsf{BF}(L, K), \rvw)\nonumber\\
&=& H(\mathsf{BF}(L, K)|\rvw) + H(\rvw).
\end{eqnarray}
Since $H(\rvw) \leq \log L$, its impact diminishes asymptotically after normalization with $L$. So we just need to consider $H(\mathsf{BF}(L, K)|\rvw)$, which can be lower bounded as
\begin{eqnarray}
&&\!\!\!\!\!\!(1/L)\!\cdot\! H(\mathsf{BF}(L, K)|\rvw)
\label{eqn:entropy-a}
\! = \!(1/L)\cdot \mathbf{E}_\rvw \log {L \choose \rvw}\\
&\geq& \!\!\!\!\!\!(1/L)\cdot \left[\min_{w: |w - (1 - p)L| \leq \epsilon L} \log {L \choose w}\right] \times\nonumber\\
&&\quad\quad \mathrm{Pr}[|\rvw - (1 - p) L|\leq \epsilon L]\nonumber\\
\label{eqn:entropy-b}
&\geq& \!\!\!\!\!\!(1/L)\!\cdot\! \log \! {L \choose (1 - p \pm \epsilon)L}\! \left[1 \!-\! 2\exp\!\left(- \frac{\epsilon^2 L^2}{2K}\right)\right]\\
&\rightarrow& \!\!\!\!\!\!h_2(p), \mbox{as}\; \epsilon \rightarrow 0 \;\mbox{and}\; L \rightarrow \infty,
\end{eqnarray}
where (\ref{eqn:entropy-a}) is due to Lemma \ref{lem:uniformity} and (\ref{eqn:entropy-b}) is due to Lemma \ref{lem:occupancy}, noting that $\lim_{L \rightarrow \infty} (1 - 1/L)^K = \exp(- \kappa)$ under the assumption of $\lim_{L \rightarrow \infty} K/L = \kappa > 0$. In an analogous way, $(1/L)\cdot H(\mathsf{BF}(L, K)|\rvw)$ can also be upper bounded by $h_2(p)$ as $L \rightarrow \infty$. Hence (\ref{eqn:bf-entropy}) is proved. The proof of (\ref{eqn:bf-entropy-conditional}) is similar and thus omitted. $\Box$

Our second result is the following:
\begin{prop}
\label{prop:ln2-achieving}
Without joint decoding, Bloom filters as channel inputs achieve the symmetric sum rate of $R_\mathrm{sum} = \ln 2$ bits/c.u., for any fixed $N$.
\end{prop}
{\it Outline of Proof:} Let each of the $N$ users have $M$ equiprobable messages, and each message be associated with a Bloom filter $\mathsf{BF}(L, K)$. The sum rate is $R_\mathrm{sum} = (N \ln M)/L$ nats/c.u.. Let $K = \kappa L/N$ for some $\kappa > 0$ which will be selected in later part of the proof.

Denote the transmitted messages of the $N$ users collectively as a length-$N$ array $\underline{\rvu}$, and the decoded messages as $\underline{\hat{\rvu}}$. The error event is $\rve = \{\underline{\hat{\rvu}} \neq \underline{\rvu}\}$.

Let the received length-$L$ array be $\underline{\rvy}$ and its weight be $\rvw$. We can lower bound the probability of correct decoding as
\begin{eqnarray}
\label{eqn:PebarLB}
&&\mathrm{Pr}[\bar{\rve}]\nonumber\\
\label{eqn:ln2-a}
&=& \sum_{w = 1}^{\min\{KN, L\}} \mathrm{Pr}[\bar{\rve}|\rvw = w] \mathrm{Pr}[\rvw = w]\\
\label{eqn:ln2-b}
&=& \sum_{w = 1}^{\min\{KN, L\}} \left[1 - \left(\frac{w}{L}\right)^K\right]^{N(M - 1)} \mathrm{Pr}[\rvw = w]\\
&>& \sum_{w/L = 1 - p - \epsilon}^{1 - p + \epsilon} \left[1 - \left(\frac{w}{L}\right)^K\right]^{N(M - 1)} \mathrm{Pr}[\rvw = w]\nonumber\\
&>& \left[1 - (1 - p + \epsilon)^K\right]^{NM} \mathrm{Pr}[|\rvw - (1 - p)L| \leq \epsilon L]\nonumber\\
\label{eqn:ln2-c}
&>& \!\!\left[1\! -\! NM(1\! -\! p + \epsilon)^K\right] \left[1\! -\! 2 \exp\left(- \frac{\epsilon^2 L^2}{2NK}\right)\right]\\
\label{eqn:ln2-d}
&>& 1 - NM(1 - p + \epsilon)^K - 2 \exp\left(- \frac{\epsilon^2 L^2}{2NK}\right),
\end{eqnarray}
with $p = \exp(-\kappa)$, where (\ref{eqn:ln2-a}) is because the number of ``1''s in $\underline{\rvy}$ is at least one and at most $\min\{KN, L\}$, (\ref{eqn:ln2-b}) is because correct decoding corresponds to that for all the $N(M - 1)$ messages which were not transmitted, their Bloom filters are not contained within $\underline{\rvy}$,\footnote{Note that this is not necessarily true for joint decoding; see the last paragraph of this section.} and (\ref{eqn:ln2-c}) is due to Lemma \ref{lem:occupancy}. For any fixed $\epsilon > 0$, the last term in (\ref{eqn:ln2-d}) is arbitrarily small as $L \rightarrow \infty$. So reliable transmission boils down to ensuring $NM(1 - p + \epsilon)^K \rightarrow 0$ as $L \rightarrow \infty$.

With $M = \exp(L R_\mathrm{sum}/N)$, we have that $NM(1 - p + \epsilon)^K \rightarrow 0$ is equivalent to
\begin{eqnarray}
N \exp\left[\frac{L}{N} \left(R_\mathrm{sum} + \kappa \ln (1 - \exp(-\kappa) + \epsilon)\right)\right] \rightarrow 0,
\end{eqnarray}
which is further equivalent to
\begin{eqnarray}
\label{eqn:RsumUB}
R_\mathrm{sum} < - \kappa \ln (1 - \exp(-\kappa) + \epsilon).
\end{eqnarray}
By letting $\epsilon \rightarrow 0$ and choosing $\kappa = \ln 2$, (\ref{eqn:RsumUB}) becomes $R_\mathrm{sum} < (\ln 2)^2$ nats/c.u., i.e., $\ln 2$ bits/c.u.. This thus establishes Prop. \ref{prop:ln2-achieving}. $\Box$

In \cite{cohen71:tcom}, \cite{hui83:dissertation} and \cite{griot06:ita}, various single-user nonlinear convolutional/trellis codes were considered, with other users' signals approximated as memoryless interference. As shown in the proof of Prop. \ref{prop:ln2-achieving}, the coding scheme based on Bloom filters does not require approximations in its performance analysis, and is valid for any fixed $N$. This result also settles an open issue in \cite{gyorfi81:it} regarding coding schemes that work for any fixed $N$; --- therein another coding scheme with random scramblers was proposed, achieving $R_\mathrm{sum} = \ln 2$ bits/c.u. only when $N$ grows exponentially with the message length.

It is interesting to note the performance gap between Prop. \ref{prop:cap-achieving} and Prop. \ref{prop:ln2-achieving}. Each user transmits a Bloom filter, and all the users' transmitted Bloom filters are superposed to form the received array. Without joint decoding, the receiver desires that for each user, exactly one of its messages is contained in the received array. With joint decoding, the receiver finds a message tuple, formed by selecting one message from every user, that exactly produces the received array; --- the receiver does allow a user to have two or more messages be contained in the received array, but may still correctly find the transmitted message by requiring each ``1'' in the received array to be contained in the Bloom filter of at least one of the transmitted messages.

\section{OR Many-access Channels}
\label{sec:bf-or-many}

We proceed to the many-access regime where the number of users, $N$, grows without bound. We assume that each user is active with probability $N_a/N$, independently with others. So the number of active users is a binomial random variable of mean $N_a$. We consider a scenario satisfying the following conditions:\\
(1) $N_a = \Theta(N^\beta)$ for some $0 < \beta < 1$;\footnote{The extreme cases of $\beta = 0$ and $1$ require a fine-grained asymptotic analysis of the proposed coding schemes and are not treated in this paper.} that is, the mean number of active users grows without bound, while the activity ratio asymptotically vanishes, with $N$.\\
(2) each user has $M$ equiprobable messages, with $M = \Theta(N^\gamma)$ for some $\gamma \geq 0$. Note that the case of a fixed number of messages corresponds to $\gamma = 0$.

First we consider the activity recognition problem. Each active user transmits a length-$L$ signature array, and each inactive user is ``silent'', i.e., transmitting a length-$L$ all-``0'' array. The receiver needs to decide, with high probability, which users are active. We characterize the efficiency of activity recognition as follows.
\begin{defn}
\label{defn:ar}
An activity recognition cost $\Omega_a$ is called feasible, if there exists a sequence of length-$(\Omega_a N_a \log_2 N)$ signature arrays such that, as $N$ grows without bound, the probability of correctly recognizing the active users converges to one.
\end{defn}

We have the following result on activity recognition.
\begin{prop}
\label{prop:ar}
The minimum feasible activity recognition cost is bounded by $1 - \beta \leq \Omega_a \leq 1/\ln 2 \approx 1.44$.
\end{prop}
{\it Outline of Proof:} The lower bound can be proved using a standard information-theoretic argument. The intuition is that by allowing all the users to fully cooperate to send a codeword informing the receiver about their activity states, the needed number of channel uses is $N h_2(N_a/N) = (1 - \beta) N_a \left[ \log_2 N + O(1) \right]$.

The upper bound is based on a specific coding scheme, using Bloom filters as signature arrays. Each user has as its signature array a Bloom filter of parameters $(L, K)$, with $K = (L/N_a) \ln 2$. An active user simply transmits its signature array, and the receiver declares the active users as those whose signature arrays as Bloom filters are contained in the received array.

Denote the activity states of the $N$ users by $\underline{\rvs}$ where $\rvs_n = 1$ if user $n$ is active and $\rvs_n = 0$ otherwise, and denote the decoded activity states by $\underline{\hat{\rvs}}$. The error event is $\rve = \{\underline{\hat{\rvs}} \neq \underline{\rvs} \}$. Note that the number of active users $\rva$ is a binomial random variable of mean $N_a$. First, we have for any $\delta > 0$,
\begin{eqnarray}
\label{eqn:ar-e-binomial}
\mathrm{Pr}[\rve] \!\!&=&\!\! \mathrm{Pr}[\rve\big | |\rva - N_a| \leq \delta N_a] \cdot \mathrm{Pr}[|\rva - N_a| \leq \delta N_a] + \nonumber\\
&& \quad \mathrm{Pr}[\rve\big | |\rva - N_a| > \delta N_a] \cdot \mathrm{Pr}[|\rva - N_a| > \delta N_a]\nonumber\\
&\leq& \!\!\!\!\!\!\max_{|a - N_a| \leq \delta N_a}\!\!\!\!\! \mathrm{Pr}[\rve\big | \rva = a] \!+\! \mathrm{Pr}[|\rva - N_a| > \delta N_a].
\end{eqnarray}
Since $\mathrm{Pr}[|\rva - N_a| > \delta N_a] \rightarrow 0$ for any $\delta > 0$ such that $\delta^2 N_a \rightarrow \infty$, we only need to ensure $\mathrm{Pr}[\rve\big | \rva = a] \rightarrow 0$ for any $(1 - \delta)N_a \leq a \leq (1 + \delta)N_a$.

Denoting the weight of $\underline{\rvy}$ by $\rvw$, we then proceed in a way similar to that in the proof of Prop. \ref{prop:ln2-achieving}, as
\begin{eqnarray}
\label{eqn:ar-bound}
&&\mathrm{Pr}[\bar{\rve}\big | \rva = a]\nonumber\\
&=& \sum_{w = 1}^{\min\{aK, L\}} \mathrm{Pr}[\bar{\rve}\big | \rva = a, \rvw = w] \mathrm{Pr}[\rvw = w \big | \rva = a]\nonumber\\
\label{eqn:ar-a}
&=& \sum_{w = 1}^{\min\{aK, L\}} \left[1 - \left(\frac{w}{L}\right)^K\right]^{N - a} \mathrm{Pr}[\rvw = w \big | \rva = a]\\
&>& \!\!\!\!\!\!\sum_{|w - (1 - p)L| \leq \epsilon L} \left[1 - \left(\frac{w}{L}\right)^K\right]^{N - a} \!\!\mathrm{Pr}[\rvw = w \big | \rva = a]\nonumber\\
&>& \!\![1 - (1 - p + \epsilon)^K]^N \mathrm{Pr}[|\rvw - (1 - p)L| \leq \epsilon L \big | \rva = a]\nonumber\\
\label{eqn:ar-b}
&>& [1 - N (1 - p + \epsilon)^K] \left[1 - 2 \exp\left(- \frac{\epsilon^2 L^2}{2 aK}\right)\right]\\
\label{eqn:ar-c}
&>& 1 - N (1 - p + \epsilon)^K - 2 \exp\left(- \frac{\epsilon^2 L^2}{2 aK}\right),
\end{eqnarray}
with $p = 2^{-a/N_a}$, where (\ref{eqn:ar-a}) is because correct activity recognition corresponds to that for all the $N - a$ inactive users, their signature arrays as Bloom filters are not contained within $\underline{\rvy}$, and (\ref{eqn:ar-b}) is due to Lemma \ref{lem:occupancy}. For any $\epsilon > 0$, the last term in (\ref{eqn:ar-c}) is arbitrarily small as $L \rightarrow \infty$. So it remains to ensure $N (1 - p + \epsilon)^K \rightarrow 0$ as $L \rightarrow \infty$, for any $(1 - \delta)N_a \leq a \leq (1 + \delta)N_a$.

Recalling that $K = (L/N_a) \ln 2$ and $L = \Omega_a N_a \log_2 N$, we have
\begin{eqnarray}
\label{eqn:ar-bound-1}
N (1 - p + \epsilon)^K &\leq& N \left[1 - \exp\left(- \frac{(1 + \delta) N_a K}{L}\right) + \epsilon \right]^K\nonumber\\
&=& N \left(1 - 2^{-(1 + \delta)} + \epsilon\right)^{\Omega_a \ln N}\nonumber\\
&=& N^{1 + \Omega_a \ln \left(1 - 2^{-(1 + \delta)} + \epsilon\right)},
\end{eqnarray}
which tends to zero for any $\Omega_a > 1/\ln 2$ by choosing sufficiently small $\delta$ and $\epsilon$. This establishes Prop. \ref{prop:ar}. $\Box$

We remark that, the activity recognition scheme also provides a non-adaptive group testing protocol \cite{chan14:it} \cite{luo08:allerton}. Extensions to noisy scenarios appear to be feasible, by slightly modifying the rule of verifying the existence of an item in a Bloom filter. An open issue is to improve the lower bound on $\Omega_a$ beyond that in Prop. \ref{prop:ar}. We also remark that, the formulation of the activity recognition problem, by allowing an asymptotically vanishing error probability rather than requiring zero error, is different from the formulation of the superimposed codes in coding theory (see, e.g., \cite{kautz64:it}).

Then we consider the message transmission problem. Each active user uniformly randomly selects a message and transmits a length-$L$ codeword array, and each inactive user is silent, i.e., transmitting a length-$L$ all-``0'' array. The receiver needs to decide, with high probability, which users are active and which messages they transmit. We characterize the efficiency of message transmission as follows.
\begin{defn}
\label{defn:mt}
A message transmission cost $\Omega_m$ is called feasible, if there exists a sequence of length-$(\Omega_m N_a \log_2 N)$ codeword arrays such that, as $N$ grows without bound, the probability of correctly recognizing the active users and decoding their transmitted messages converges to one.
\end{defn}

We have the following result on message transmission.
\begin{prop}
\label{prop:mt}
The minimum feasible message transmission cost is bounded by $1 - \beta + \gamma \leq \Omega_m \leq (1 + \gamma)/\ln 2$.
\end{prop}
{\it Outline of Proof:} The lower bound can be proved using a standard information-theoretic argument. The intuition similar to that of Prop. \ref{prop:ar} is that, by allowing all the users to fully cooperate to send a codeword informing the receiver about the messages of active users, the needed number of channel uses is $N h_2(N_a/N) + N_a \log_2 M = (1 - \beta + \gamma) N_a [\log_2 N + O(1)]$.

The upper bound is based on a specific coding scheme which consists of two phases. The new idea different from existing works (e.g., \cite{chen16:arxiv}) is the following which we call partial activity recognition: Phase 1 need not be long enough to ensure accurate activity recognition, but instead, the receiver makes up a list of believed active users that are roughly twice as many as truly active users; Phase 2 then resolves this ambiguity along with decoding the messages. An error occurs if either an active user has at least a message which is not transmitted but is falsely contained in the received array in Phase 2, or an inactive user is falsely recognized as active in Phase 1 and has at least a message falsely contained in the received array in Phase 2.

Let the Bloom filters in Phase $i$ be of parameters $(L_i = \kappa_i N_a \log_2 N, K_i = (L_i/N_a)\ln 2)$, $i = 1, 2$. Similar to (\ref{eqn:ar-e-binomial}), the probability of the number of active users $\rva$ significantly deviating from its mean asymptotically vanishes with $N$ and we only need to ensure $\mathrm{Pr}[\rve | \rva = a] \rightarrow 0$ for any $(1 - \delta)N_a \leq a \leq (1 + \delta)N_a$, for sufficiently small $\delta > 0$.

Denote the received array in Phase $i$ by $\underline{\rvy}_i$, and its weight by $\rvw_i$, $i = 1, 2$. We have
\begin{eqnarray}
&&\!\!\!\!\!\!\!\!\mathrm{Pr}[\bar{\rve} \big | \rva = a] = \sum_{w_1, w_2} \mathrm{Pr}[\bar{\rve} \big | \rva = a, \rvw_1 = w_1, \rvw_2 = w_2]\times\nonumber\\
&&\quad\quad\quad \mathrm{Pr}[\rvw_1 = w_1, \rvw_2 = w_2 \big | \rva = a].
\end{eqnarray}
Denoting $\mathrm{Pr}[\bar{\rve} \big | \rva = a, \rvw_1 = w_1, \rvw_2 = w_2]$ by $q(a, w_1, w_2)$, we have
\begin{eqnarray}
&&\mathrm{Pr}[\bar{\rve} \big | \rva = a] = \sum_{w_1, w_2} q(a, w_1, w_2)\times\nonumber\\
&&\quad\quad \mathrm{Pr}[\rvw_1 = w_1 \big | \rva = a] \mathrm{Pr}[\rvw_2 = w_2 \big | \rva = a]\nonumber\\
&>& \min_{|w_i - (1 - p)L_i|\leq \epsilon L_i, i = 1, 2} q(a, w_1, w_2) \times\nonumber\\
&&\quad \mathrm{Pr}[|\rvw_1 - (1 - p)L_1| \leq \epsilon L_1 \big | \rva = a]\times\nonumber\\
&&\quad \mathrm{Pr}[|\rvw_2 - (1 - p)L_2| \leq \epsilon L_2 \big | \rva = a]\nonumber\\
&>& \min_{|w_i - (1 - p)L_i|\leq \epsilon L_i, i = 1, 2} q(a, w_1, w_2) \times\nonumber\\
&&\!\!\!\!\!\!\!\!\!\! \left[1 - 2\exp\left(-\frac{\epsilon^2 L_1^2}{2aK_1}\right)\right]\!\! \left[1 - 2\exp\left(-\frac{\epsilon^2 L_2^2}{2aK_2}\right)\right],
\end{eqnarray}
where $p = 2^{-a/N_a}$. So it remains to ensure, for sufficiently small $\epsilon > 0$, $\min_{|w_i - (1 - p)L_i|\leq \epsilon L_i, i = 1, 2} q(a, w_1, w_2) \rightarrow 1$ as $N \rightarrow \infty$, for any $(1 - \delta)N_a \leq a \leq (1 + \delta)N_a$. For this, we note that
\begin{eqnarray}
\label{eqn:q}
&&q(a, w_1, w_2) = \left[1 - \left(\frac{w_2}{L_2}\right)^{K_2}\right]^{a(M - 1)}\times \nonumber\\
&&\!\!\!\!\!\!\!\!\left\{1 - \left(\frac{w_1}{L_1}\right)^{K_1} \left[1 - \left[1 - \left(\frac{w_2}{L_2}\right)^{K_2}\right]^{M}\right]\right\}^{N - a}\!\!,
\end{eqnarray}
where the first term corresponds to the probability that none of the active users has its message falsely decoded, and the second term corresponds to the probability that none of the inactive users is falsely recognized as active and has any of its messages falsely decoded. After manipulations of (\ref{eqn:q}), we find that it suffices to have
\begin{eqnarray}
\label{eqn:kappa-1-2-bounds}
\kappa_2 \ln 2 - \beta - \gamma > 0, \;\; (\kappa_1 + \kappa_2) \ln 2 - 1 - \gamma > 0;
\end{eqnarray}
that is, we can choose any $\kappa_1 > (1 - \beta)/\ln 2$, $\kappa_2 > (\beta + \gamma)/\ln 2$, and sufficiently small $\delta$ and $\epsilon$, to ensure $\mathrm{Pr}[\rve] \rightarrow 0$ as $N \rightarrow \infty$. In total, the two-phase coding scheme requires $(\kappa_1 + \kappa_2) N_a \log_2 N$ c.u.s, where $\kappa_1 + \kappa_2$ can be any number greater than $(1 + \gamma)/\ln 2$. This establishes Prop. \ref{prop:mt}. $\Box$

According to the coding scheme in Prop. \ref{prop:ar}, accurate activity recognition needs $\kappa_1 > 1/\ln 2$, stricter than $\kappa_1 > (1 - \beta)/\ln 2$ in (\ref{eqn:kappa-1-2-bounds}). Nevertheless, due to the gap between the lower and upper bounds on $\Omega_a$, at this point we cannot affirmatively assert that partial activity recognition is indeed optimal. This is an open issue for further research.

When $\gamma = 0$, each user has a fixed number of messages, and the bounds in Prop. \ref{prop:ar} and Prop. \ref{prop:mt} coincide, i.e., the cost of activity recognition dominates.

We characterize the complexity of our coding schemes in terms of the average number of hash functions needed for accomplishing encoding or decoding. Our result is as follows.
\begin{prop}
\label{prop:complexity}
For the coding schemes in the proofs of Prop. \ref{prop:ar} and Prop. \ref{prop:mt}:\\
(1) Each active user needs to hash $O(\ln N)$ times for encoding its signature/codeword array.\\
(2) For activity recognition, the receiver needs to hash, on average, $O(1)$ times per user.\\
(3) For message transmission, the receiver needs to hash, on average, $O(\max\{1, N^{\beta + \gamma - 1}\})$ times per user.
\end{prop}
{\it Outline of Proof:} Result (1) follows from the fact that the Bloom filters in the coding schemes in the proofs of Prop. \ref{prop:ar} and Prop. \ref{prop:mt} are all of parameters $\left(O(N_a \ln N), O(\ln N)\right)$. For proving result (2), note that for verifying an inactive user's Bloom filter signature array, as soon as a hashed position in the received array is ``0'', the receiver can discard this inactive user early, incurring only $O(1)$ hashes on average. Furthermore, the ratio between the mean number of active users and the total number of users is asymptotically vanishing. Result (3) follows analogously and the details of derivation are thus omitted. $\Box$

The complexity result for message transmission exhibits a threshold behavior. When $\beta + \gamma > 1$, roughly corresponding to $N_a M \gg N$, the decoding complexity per user grows unbounded with $N$. Otherwise, the decoding complexity per user is bounded.

\section*{Acknowledgement}

The authors are grateful to Dongning Guo for valuable discussions and sharing an early version of \cite{chen16:arxiv}. The work has been supported in part by the National Natural Science Foundation of China under Grant 61379003.


\end{document}